# Using Lightweight Activity Diagrams for Modeling and Generation of Web Information Systems

Dirk Reiß[1] and Bernhard Rumpe[2]

[1] Institute for Building Services and Energy Design
Technical University of Braunschweig, Germany
http://www.igs.bau.tu-bs.de
[2] Software Engineering
RWTH Aachen University, Germany
http://www.se-rwth.de

**Abstract.** The development process of web information systems nowadays improved a lot regarding effectiveness and tool support, but still contains many redundant steps for similar tasks. In order to overcome this, we use a model-driven approach to specify a web information system in an agile way and generate a full-fledged and runnable application from a set of models. The covered aspects of the system comprise data structure, page structure including view on data, page- and workflow within the system as well as overall application structure and user rights management. Appropriate tooling allows transforming these models to complete systems and thus gives us opportunity for a lightweight development process based on models. In this paper, we describe how we approach the page- and workflow aspect by using activity diagrams as part of the agile modeling approach MontiWIS. We give an overview of the defined syntax, describe the supported forms of action contents and finally explain how the behavior is realized in the generated application.

## 1 Introduction

Even though the development of web information systems is supported by a variety of web frameworks in almost all modern programming languages (see [1] for an overview of those), it still requires a lot of repetitive, tedious and error prone work for rather basic tasks like the definition of data structure and the creation of CRUD (create, read, update and delete) functionality with according pages therefor or basic but system-wide consistent user and rights management. To alleviate the effort to implement a web information system, we developed the model-driven approach MontiWIS to abstract from the details of the repetitive implementation tasks while maintaining flexibility where necessary. Following an agile development approach, we allow rapid prototyping by providing extensive default behavior for a minimal set of models and stepwise refinement of the applications' functionality through the addition of further models and aspects to the system.

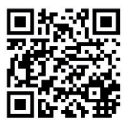

As the data model is the basic concept in our approach, a developer can start his application model by only specifying this part of the system. The generation process results in a full-fledged runnable system with default CRUD functionality for the given data model. From this point, further aspects like user- and rights management or special views on the data model which in turn can be incorporated in complex business processes can be specified. As processes and business logic are a crucial part of a web information system, we focus on this aspect in the following and describe the different development stages from the specification using a profile of UML activity diagrams [2] to the technical realization in the generated application.

The paper is structured as follows. In Section 2, we shortly describe the MontiWIS approach in a whole and explain the interdependency of the different languages and aspects of the system. Section 3 introduces a graphical example, describes its equivalent in the textual syntax of activity diagrams and explains the supported action contents. In Section 4, we give an overview of the technical realization of processes in the generated system. In Section 5, similar approaches are discussed and Section 6 concludes the paper and gives an outlook on planned extensions of our approach.

## 2 Overview of MontiWIS Development Method

MontiWIS is the successor of MontiWeb [3] and extends it especially in the area of user- and rights management, application construction and page description expressivity. Both approaches use MontiCore [4, 5] and its infrastructure to define the abstract and concrete syntax of their textual languages and to process the models.

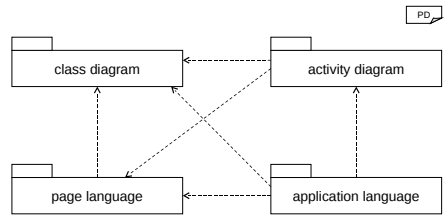

**Fig. 1.** Languages and their relation in MontiWIS

The package diagram in Figure 1 shows the different models that are used to specify structure and behavior of the application. Each package in the figure depicts one modeling language that covers a certain aspect of the system. An arrow from one language to another indicates that parts of the targeted language are referenced by the source language. The UML/P [6, 7] class diagrams, which are used to describe the data structure of the web information system, are based on the grammar and infrastructure of the work by Martin Schindler [8]. As an addition to his implementation, we offer domain specific types like for instance String, Email, Date and Text that result in the generation of type-specific input elements with consistency checks and a reasonable default behavior. The domain specific page language facilitates the description of a user interface page with elements like, among others, headings, simple text parts and complex tables. Pages may have input parameters (whose types are usually classes from the class diagram) that can be displayed in a whole or attribute-by-attribute in either an editable (i.e. using input forms) or non-editable way (by just outputting their values). The activity diagrams are used to describe the control- and data flow between pages as well as complex multi-user workflows. An action, being the executable part of an activity, may contain references to a page that will be displayed when an action is executed. It may also reference classes and operate on them as part of the included business logic. This way, we support both, actions that require user interaction through page display as well as completely automatic ones that are executed on the server in a standalone manner. However, in most cases, code is executed before and after the presentation of a page. Section 3 describes the language in detail. Finally, the domain specific application language serves as central model to define flexible roles and associated rights (e.g., guest, student or administrator) in the application. It allows the definition of the systems' navigation menu and, considering different roles, different views on the application as a whole. Elements of all three languages can be referenced and included as menu entries or can have rights restrictions applied to them. Classes are referenced implicitly by including default pages for creating new or listing all objects of a certain type, static pages can be included to display predefined content and activities are referenced as workflows that can be invoked.

By using different languages for different aspects of the system, we can modify one aspect (such as user roles or page structure) without the need to adapt related others (e.g., activities that may be invoked by a certain role or ones that use the page). Although all these languages are defined independently, the MontiCore infrastructure easily allows us to define and check connections between these languages. This way we can for instance check the existence of classes referenced as data types in either pages or activities, check whether attributes are used correctly or ensure that pages and activities are linked correctly in menu and rights declarations.

As our lightweight approach is designed with agile methodologies in mind, reasonable default behavior is generated in the final application where specific details are omitted. This approach allows for early results while maintaining flexibility where needed.

Currently, the MontiWIS generator creates a full-fledged Java EE application that uses JPA for persistence mechanisms and JSP pages (with included JQuery on the client side) to create the modular and AJAX-enabled user interface of the application.

## 3 Page- and Workflow using Activity Diagrams

In order to explain the principles and possibilities of activity diagrams in MontiWIS, we first introduce the example given in Figure 2. Afterwards, we explain how the described example can be specified by using the language and constructs of our approach.

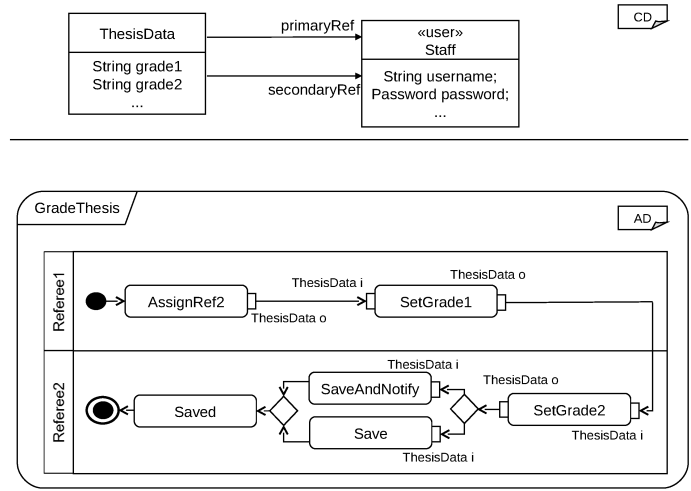

**Fig. 2.** Example: class and activity diagram to grade a thesis

The diagrams show an excerpt of a system that supports a lecturer in his curriculum tasks. The relevant part of the underlying data structure is given in the upper half. It consists of only two classes, **ThesisData** and **Staff**, each with just a subset of the attributes included here. Besides two attributes that hold grades (**grade1** and **grade2**), **ThesisData** includes two associations to **Staff** that represent the thesis' referees (**primaryRef** and **secondaryRef**). **Staff** is stereotyped with **user**, which indicates that this domain model class also holds information to log in to the system and that objects of this class can serve as actors in a workflow.

The activity diagram in the lower part of the figure shows the workflow that guides two different participants through the process of assigning grade values to a thesis. As **Referee1** is the partition that contains the first action in the workflow, the person starting the activity is automatically assigned to it. In the first action of the activity (**AssignRef2**), one **Staff** is selected from a list of all possible ones and then assigned to partition **Referee2**. In order to complete the **ThesisData** object passed on along the workflow (indicated by pin o), both, the **Staff** actually executing the workflow and the one selected within the action, will need to be set as well. Afterwards, the **ThesisData** object is handed over to the next action **SetGrade1**, also conducted in the same partition and thus by the same user. In this action, the user determines and assigns a value to attribute `grade1` in the given object. After this step, no further tasks need to be executed by the first referee and the `ThesisData` object is passed on to action `SetGrade2` that is associated with the other partition. Therefore, this workflow will appear in the list of active workflows of the user previously associated with this partition. Just as the `Referee1`, `Referee2` also needs to assign his grade to the thesis data, which he is asked to do in action `SetGrade2` when he resumes the activity. At this point, the workflow is split up in two alternative directions, either of which he can take. As there are no guards that could be evaluated automatically, the user will be prompted for the desired direction. In our approach, each possibility is represented by a button that is included in the page at execution time. The subsequent action `SaveAndNotify` persists the object in the database and sends a notification to all participants while `Save` only persists the object. After either of these actions, the action `Saved` notifies the user of the successful action by displaying an appropriate page. A final submission of this page finishes the workflow and cleans up runtime data.

As already mentioned in Section 2, we use a textual syntax to describe the system. This also applies to the models described above. For the sake of brevity and focus of this paper, we omit the textual representation of the given class diagram and only describe (parts of) the activity diagram language in Figure 3.

The diagram itself is identified by the keyword `activity` and a unique name (1). A partition of an activity is defined by the keyword `role`, its name and a set of actions that are associated with them in curly braces (3). An action is denoted by the keyword `action` and a name that has to be unique within the activity. The actions' content is included in braces (6-20). In case an action contains in- or output parameters (or both), we declare them at the beginning of the action (7, 25). The keywords `in` and `out` denote the type of the parameters that are listed after a colon and separated by a comma. A parameter declaration consists of its type (which is usually defined in the class diagram or one of a few allowed Java data types such as `Set` (8)) and a name that is scoped within the action and thus can be used throughout its contents. The same applies to variables (denoted by the keyword `var` (8)), which can be used within the action but are not accessible from outside of it.

The business logic within an action is executed in the order it is defined. Here, we first invoke several predefined commands. A command consists of the keyword `cmd` and a colon, followed by the syntactically predefined command itself. These commands are specific to the domain of web information systems and provide functionality such as database access, or mechanisms for workflow control. The command in line 10 for instance retrieves a set of all `Staff` objects from the database and assigns them to the previously declared variable `allStaff`. The command in line 11 allocates the variable `actualUser` with the actually logged in user object (which is of type `Staff`), which then can be accessed in the remainder of the action.

The keyword `view` (12) indicates the presentation of a page as part of the actions' logic. The page `SelectSecondaryRef` is defined in a separate model and

───────────────── Activity diagram ─────────────────
```
1    activity GradeThesis {
2
3      role Referee1 { AssignRef2, SetGrade1, ... }
4      // ...
5
6      action AssignRef2 {
7        out : ThesisData o;
8        var : Set<Staff> allStaff, Staff actualUser, Staff selectedUser;
9
10       cmd : allStaff = Staff.loadAll();
11       cmd : actualUser = getActualUser();
12       view : SelectSecondaryRef(allStaff);
13       java : {
14         selectedUser = allStaff.iterator().next();
15         o = new ThesisData();
16         o.setPrimaryRef(actualUser);
17         o.setSecondaryRef(selectedUser);
18       }
19       cmd : assignRole(Referee2, selectedUser)
20     }
21
22     action SetGrade1 { ... }
23
24     action SetGrade2 {
25       in : ThesisData i;
26       out : ThesisData o;
27
28       view : SetGrade2Page(i);
29       java : {
30         o = i;
31       }
32     }
33
34     // other actions omitted here
35
36     initial -> AssignRef2;
37     AssignRef2.o -> SetGrade1.i;
38     SetGrade2.o -> SaveAndNotify.i | Save.i;
39     // other edges omitted here
40   }
```
────────────────────────────────────────────────────

**Fig. 3.** Activity diagram in textual syntax

accepts a set of `Staff` objects as parameter. Pages generally accept one or more objects as input and return them with updated values after submission of the page. In this case, the page contains a listing view of the given set of objects and allows the selection of a single object. Therefore, the variable `allStaff` contains a single `Staff` element after line 12.

In order to allow as much flexibility in declaring business logic as possible, we allow plain Java code to be included in an action. A block of code begins with the keyword `java`, followed by a colon and surrounded by braces (13-18). Inside them, arbitrary code may be included. Here, we first retrieve the selected `Staff` object from the submitted set of objects (14), followed by the instantiation of a new `ThesisData` object (15) and the assignment of both, the actual and the manually selected `Staff` to the domain object (16-17). Finally, the selected `Staff` object is assigned to the role named `Referee2` (19).

The action `SetGrade2` (24-32) contains in- and output parameters and will display the object assigned to the input parameter through page `SetGrade2Page` (26). This page is defined in a way that only one of the two grade attributes is editable and thus only the correct attribute can be updated. Finally, the submitted `ThesisData` is assigned to the output parameter (30). As described in the example before, the action `SetGrade1` facilitates a similar behavior, just different parts of the object are editable (by a different user). The remaining actions are omitted here due to lack of space.

Lines 36-38 show some of the possible ways to define activity edges. An edge is defined by a source node (and optional output parameter, separated by a dot) on the left hand side of an arrow (`->`) and the target node (and possible input parameter) on its right hand side, all identified by their node and parameter names. Line 36 defines the edge from the activities' initial node (denoted by the keyword `initial`) to action `AssignRef2`. Line 37 defines an edge from the output parameter of action `AssignRef2` to the input parameter of action `SetGrade1`. Line 38 shows an abbreviated syntax for a decision node (indicated by the pipe symbol (`|`)). Here, the output parameter of `SetGrade2` is connected with the input parameter of either `SaveAndNotify` or `Save`.

This activity diagram demonstrates a subset of the features that we offer to model business logic in our web information system. It provides both, flexibility by including Java code directly but also domain specific functionality like the presentation of pages, access to database and server runtime information as well as activity execution information. When designing the language, we focused on compact syntax and preferred to include pure Java syntax to defining a separate action language that would cover the same functional range. As we actually focus on Java as a target platform for generated code, we can easily include existing functionality from, e.g., external APIs or manually implemented complex logic almost instantly by invoking them from within an action. Nevertheless, other languages (like, e.g., C#, PHP or an abstract action language as used in executable UML [9]) could be included. Furthermore, the used Java code could as well be used to generate code in the above-mentioned languages, which would require a more complex code generator and appropriate mappings from Java constructs to corresponding ones in the desired target language.

## 4 Technical Realization

After describing how workflows and included business logic are specified in MontiWIS, we now focus on the technical realization of these aspects in the generated system. The execution follows the semantics we defined in [10] and thus the behavior given by the UML itself. As we wanted control over the way we execute both, simple page- and complex workflow, we preferred to implement the execution components ourselves instead of using freely available workflow engines such as Activiti [11] or JBPM [12] whose integration needed rather complex changes to allow the behavior we envisioned. A future integration of such existing engines is possible but for the time being not planned. Figure 4 shows a simplified overview of the architecture used to process and execute the workflows.

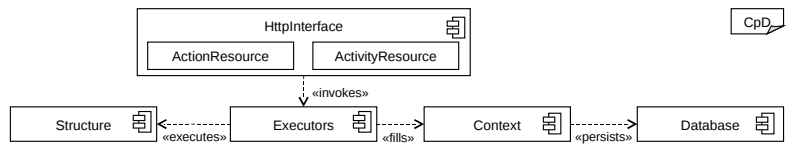

**Fig. 4.** Execution architecture of activities

As a basic principle, MontiWIS follows a REST [13] approach when interacting with the application. This implies that basic constructs of the application are treated as resources that are addressed by URLs and manipulated by HTTP requests. Besides the business domain objects (specified by the class diagram), we also treat activities and actions as resources. The two classes ActionResource and ActivityResource (part of the component HttpInterface) are accepting requests to both, actions and activities. The former is used to interact with a workflow once it has been started, the latter provides access to activities that need to be started, stopped or continued. The component Structure consists of the classes that are generated from the input models. They contain model-specific information such as activity structure or business logic in the case of an action. The component Context holds information about execution state such as actual values of node parameters or some sort of token representation to reflect the overall activity state. These classes are persisted in a database and thus can be retrieved at a later point to continue system execution or to share process state among different users of the application. The Executor component merges the runtime information from the contexts with the structure of the actual activities. Its classes include logic to control activity execution and update state, always specific to the type of the node. Each of the components contains classes with node-specific functionality for different types of nodes in an activity, such as for instance actions or decision nodes.

In order to explain typical steps of activity execution, we illustrate the different requests and responses occurring by means of the actions of Referee2 in

Figure 2. The sequence diagram in Figure 5 shows HTTP requests and how they are processed in our application.

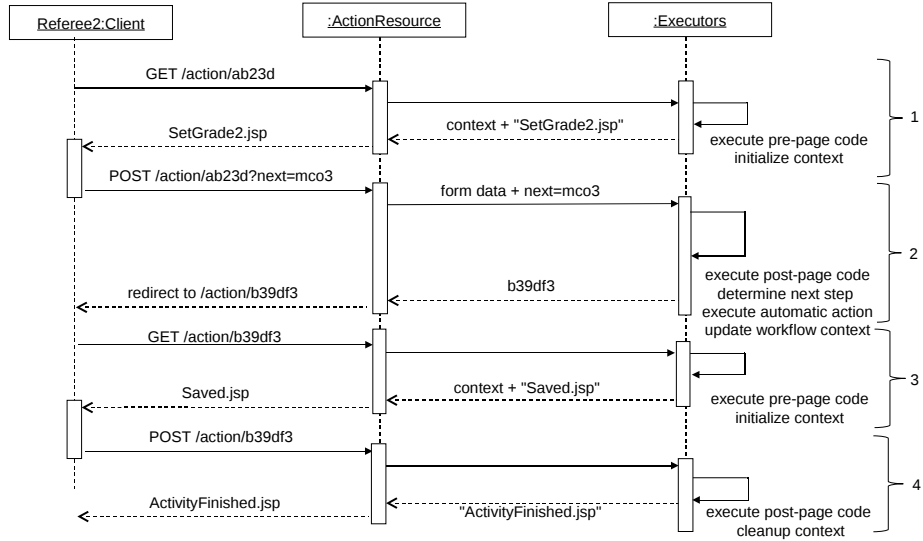

**Fig. 5.** Sequence diagram of HTTP interaction with workflows

The participating actors in the sequence diagram are the Client that interacts with the system through a web browser, the ActionRessource (which is invoked through an URL that addresses actions through the part /action as depicted in the figure) that receives the HTTP requests and returns the response to the client. The last actor (labeled Executors for the sake of simplicity) represents the different executor classes involved in the actual processing.

As we mentioned earlier, the execution of an action is usually split into a block of code executed before the display of the page, the page presentation itself and the code executed after the submission of possible form data. Step 1 in Figure 5 is initiated by a GET request to an action identified by the appended id (ab23d in this case). The resource class calls the executors to process the code defined before the page presentation and to initialize the page context, which contains all necessary information such as the ThesisData to display. Both, the context and the SetGrade2 JSP page itself are returned to the resource class and, after the execution of the JSP code, sent back to the client. As described in Section 3, decision nodes that do not contain any guards are interpreted as a free choice for the user of the application. Thus, the generated and presented page will contain elements that let the user choose which path to take after page submission. As shown in step 2, additionally to the form data, information about the users' decision is also sent to the resource class. The form data as well as the chosen path are passed to the executors that process potential post-page code, determine the next page based on the user decision and update the activities runtime context. As the next action is a completely automatic one that does not contain any user interaction whatsoever, the adequate logic will be executed, which is also part of the executors in step 2. Afterwards, the activity state is updated again and the identifier of the next action that also contains user interaction is returned to the ActionResource class. The identifier is used to address the next action and a HTTP redirect is send to the user that points to this action. In step 3, a GET request addressing the redirection target is automatically issued by the users' browser, the pre-page code is executed and a page indicating a successful save operation is displayed. Although the next step in the activity is the final node, we still demand another page submission in order to trigger possible post-page code to be executed. This call is depicted in step 4. Finally, the execution context is cleaned up and removed from the database.

## 5 Related Work

In the context of model driven development of web information systems, several approaches exist that also offer support for page- and workflow, most of them exploiting graphical models to describe the system. An evaluation of different web modeling approaches can be found in [14].

WebML [15] offers workflow concepts by incorporating process execution elements in their global model and, in contrast to our approach, do not offer a separate kind of model. Business logic is basically specified through a set of parametrized operation elements which does not offer the flexibility of plain Java code as MontiWIS allows.

WebWorkFlow [16] employs a textual domain specific language to describe processes with different aspects like, e.g., user access control or page references included in the process model. Here, model transformation approaches are used to transform higher level constructs to the more basic ones of WebDSL [17]. Unlike this approach, we do not use the basic concepts of some underlying domain specific language but rather our executor classes to enforce the process execution semantics.

The UWE approach [18] and especially its application UWE4JSF [19] are based on pure UML profiles. This approach uses annotated and stereotyped class diagrams to describe basic navigation and, although activity diagrams are also allowed to specify process logic, does not offer rich support for distributed processes. Furthermore, it focuses on atomic method calls and does not provide support for a close integration of user interaction and business logic.

The effort to integrate process execution into OOWS [20] uses BPMN as notation and model transformations to generate an extended OOWS navigation model (which itself is transformed to web pages) as well as WS-BPEL code to execute the process logic externally. Thus, in contrast to our approach, the process is not an integral part of the system but runs outside of the normal application logic.

OOHDM [21] offers means to tightly integrate processes and business logic in the navigational and conceptual models and does not allow separate models as MontiWIS does. Furthermore, complex business logic is rather hard to integrate, as interfaces to external program resources are not easily accessible.

Most of these approaches require a complete set of models covering all aspects of the system to generate the application and, unlike MontiWIS, do not provide extensive default behavior where parts of the application models are missing.

## 6 Conclusion and Future Work

In this paper, we described the business logic and workflow aspect of our modeling approach to alleviate the effort to develop web information systems. We shortly gave an overview of the basic concepts of our approach and how they are interrelated. We then introduced an illustrating example and gave a subset of the textual notation that we use in our system, highlighted how we interpret the execution of actions and activities and what possibilities we offer to specify action contents. Afterwards, we gave a simplified overview of our server side architecture regarding activity execution and described how typical workflow execution is carried out between a client and our server components.

Although the actual state of our implementation gives us great flexibility to express almost arbitrary business logic, we still see some areas of improvement. One actual focus of our work is the usability of our approach. This encompasses the development of workflow libraries that exploit hierarchical decomposition of activities to extract components that can be reused in other contexts. For the same reason, we plan to implement more commands to offer convenience functionality such as email transport or more fine-grained database search capabilities. Such functionality can actually be incorporated through plain Java calls but as our experience shows, could use closer integration with the generated system and runtime environment.

12      Reiß and Rumpe